\documentstyle[epsfig,psfig]{style}

\begin{document}
\title{Oscillation and Instabilities of Relativistic Stars
\footnote{in ''Recent Developments in General Relativity'',
Proceedings of XIV Conference of General Relativity and
Gravitational Physics, SIGRAV-2000, Genoa, Springer-Verlag (2001)
Ed. R. Cianci }}

\author{ Kostas D. Kokkotas $^{1}$ and Nils Andersson $^{2}$ }

\address{(1) Department of Physics, Aristotle University of Thessaloniki,
Thessaloniki 54006, Greece \\
}
\address{(2) Department of Mathematics, University of Southampton,
Southampton SO17 1BJ, U.K.
}


\begin{abstract}
In this short review we discuss the relevance of
ongoing research into
stellar oscillations and associated instabilities
for the detection of gravitational waves and the future
field of ``gravitational-wave astronomy''.
\end{abstract}

\section{Introduction}

As we enter the new millennium there is a focused worldwide effort
to construct devices that will enable the first undisputed
detection of gravitational waves. A network of large-scale
ground-based laser-interferometer detectors (LIGO, VIRGO, GEO600,
TAMA300) is due to come on-line soon, and the sensitivity of the
several resonant mass detectors that are already in operation
continues to be improved. An integral part in this effort is
played by theoretical modeling of the expected sources. Theorists
are presently racking their brains to think of various sources of
gravitational waves that may be observable once the new
ultra-sensitive detectors operate at their optimum level, and of
any piece of information one may be able to extract from such
observations.

The theory of stellar pulsation
is richly endowed with interesting phenomena, and ever
improving observations suggest that most stars exhibit complicated
modes of oscillation. Thus it is natural to try to match
theoretical models to observed data in order to extract information  about
the dynamics of distant stars. This interplay
between observations and stellar pulsation theory is known as
asteroseismology.

One of the most challenging goals that can (at least, in principle)
be achieved via gravitational-wave detection is the determination of
the equation of state of matter at supranuclear densities.
We have recently argued that observed gravitational waves from
the various nonradial pulsation modes of a neutron star can be
used to infer both the mass and the radius of the star with
surprisingly good accuracy, and thus put useful constraints on the
equation of state \cite{ak96,ak98,KAA01,BBF99,Yip99}.
But it is not clear that
astrophysical mechanisms can excite the various oscillation
modes to a detectable level. It seems likely that only the most violent
processes, such as the actual formation of a neutron star following a supernova or
a dramatic starquake following, for example, an internal phase-transition,
will be of relevance. The strengthening evidence for
magnetars \cite{duncan}, in which a
starquake could release large amounts of energy, is
also very interesting in this respect. One can estimate that
these events must take place in our immediate neighborhood
(the Milky Way or the Local Group) in order to be observable.

The spectrum of a pulsating relativistic star is known to be
tremendously rich, but most of the associated pulsation modes
are of little relevance for gravitational-wave detection.
From the gravitational-wave point of view one would expect
the most important modes (in addition to the unstable $r$-modes
\cite{ak01})
to be the fundamental ($f$) mode of fluid oscillation,
the first few pressure ($p$) modes
and the first gravitational-wave ($w$) modes \cite{ks92}.
For more details on the theory of relativistic
stellar pulsation we refer the reader to two recent review articles
 \cite{kokkotas97,kslrev}. That the bulk of the energy from an
oscillating neutron star is, indeed, radiated through these modes has
been demonstrated by numerical experiments
\cite{aaks,2ns,ruoff,Tomina99,AP99,ferkdk,ruoff1}.

Consequently it is relevant to try to device a strategy for
detecting gravitational waves from pulsating neutron stars, and
see what information such a detection could provide regarding the
stars parameters. Such a strategy can potentially be of great
importance to gravitational-wave astronomy, since most stars are
expected to oscillate nonradially. In principle, one would expect
the modes of a star to be excited in any dynamical scenario that
leads to significant asymmetries.  Of course, one can only hope to
observe gravitational waves from the most compact stars (neutron
stars and possibly strange stars) and only the most violent
processes are of interest. Still, there are several scenarios in
which the various pulsation modes may be excited to an interesting
level: (a) A supernova explosion is expected to form a wildly
pulsating neutron star that emits gravitational waves. The current
estimates for the energy radiated as gravitational waves from
supernovae is rather pessimistic, suggesting a total release of
the equivalent to $10^{-6} M_\odot c^2$, or so. However, this may
be a serious underestimate if the gravitational collapse in which
the neutron star is formed is strongly non-spherical. Optimistic
estimates suggest that as much as $10^{-2}M_\odot c^2$ may be
released in extreme events. (b) Another potential excitation
mechanism for stellar pulsation is a starquake, e.g., associated
with a pulsar glitch. The typical energy released in this process
may be of the order of the maximum mechanical energy that can be
stored in the crust, estimated to be at the level of
$10^{-9}-10^{-7}M_\odot c^2$ \cite{blaes,mock}. This is also an
interesting possibility  considering the recent suggestion that
the soft-gamma repeaters are magnetars,  neutron stars with
extreme magnetic fields \cite{dt92}, that undergo frequent
starquakes. It seems very likely that some pulsation modes are
excited by the rather dramatic events that lead to the most
energetic bursts seen from these systems. Indeed, Duncan
\cite{duncan} has recently argued that toroidal modes in the crust
should be excited. If modes are excited in these systems, an
indication of the energy released in the most powerful bursts is
the  $10^{-9}M_\odot c^2$  estimated for the March 5 1979 burst in
SGR~0526-66. The maximum energy should certainly not exceed the
total supply in the magnetic field $\sim 10^{-6} (B/10^{15} G)^2
M_\odot c^2$ \cite{dt92}. The possibility that a burst from a soft
gamma-ray repeater may have a gravitational-wave analogue is very
exciting. (c) The coalescence of two neutron stars at the end of
binary inspiral may form a pulsating remnant. It is, of course,
most likely, that a black hole is formed when two neutron stars
coalesce, but even in that case the eventual collapse may be
halted long enough (many dynamical timescales) that several
oscillation modes could potentially be identified \cite{baum}.
Also, stellar oscillations can be excited by the tidal fields of
the two stars during the inspiral phase that precedes the merger
\cite{KS95}.   (d) The star may undergo a dramatic
phase-transition that leads to a mini-collapse. This would be the
result of a sudden softening of the equation of state (for
example, associated with the formation of a  condensate consisting
of pions or kaons). A phase-transition could lead to a sudden
contraction during which a considerable part of the stars
gravitational binding energy would be released, and it seems
inevitable that part of this energy would be channeled into
pulsations of the remnant. Large amounts of energy that could be
released in the most extreme of these scenarios: a contraction of
(say) 10\% can easily lead to the release of $10^{-2}M_\odot c^2$.
Transformation of a neutron star into a strange star is likely to
induce pulsations in a similar fashion. It is reasonable to assume
that the bulk of the total energy of the oscillation is released
through a few of the stars quadrupole pulsation modes in all of
these scenarios. We will assume that this is the case and assess
the likelihood that the associated gravitational waves will be
detected. Having done this we discuss the inverse problem, and
investigate how accurately the neutron star parameters can be
inferred from the gravitational wave data. Finally, we  briefly
discuss the gravitational-wave driven instability of the so-called
$r$-modes.

\section{Nonradial stellar oscillations: Theoretical minimum}

A neutron star has a large number of families of pulsation modes
with more or less distinct character.
For the simplest stellar models, the relevant modes are
high frequency pressure $p$-modes and the low frequency
gravity $g$-modes \cite{unno}.  For a typical nonrotating neutron star
model the fundamental $p$-mode (usually referred to as the $f$-mode),
whose eigenfunction has no nodes
in the star,
has frequency in the range 2-4~kHz, while the
first overtone lies above $4$~kHz. The $g$-modes depend sensitively on
the internal composition and temperature distribution, but they typically
have frequencies of a few hundred Hz.
The standard mode-classification
dates back to the seminal work of Cowling \cite{cow41}, and is based
on identifying the main restoring
force that influences the fluid motion.
As the stellar model is made more detailed and
further restoring forces are included
new families of modes come into play. For example, a neutron star model
with a sizeable solid crust separating a thin ocean from a central
fluid region will have $g$-modes associated with both the
core and the ocean as well as modes associated with
shearing motion in the crust\cite{mcdermott,strohmayer}.
Of particular interest to relativists is the existence of a class of
modes uniquely associated with the spacetime itself
\cite{ks86,kojima88,ks92,akk96};
the so-called
$w$-modes (for gravitational \underline{w}ave). These modes
essentially arise because the curvature
of spacetime that is generated by the background density distribution
can temporarily trap impinging gravitational waves. The $w$-modes typically
have high frequencies (above 6~kHz) and damp out in a fraction of a
millisecond. It is not yet clear whether one should expect these modes
to be excited to an appreciable level during (say) a gravitational
collapse following a supernova. One might argue that they
provide a natural channel for the release of any initial
deformation of the spacetime, but there are as yet no solid evidence
indicating a significant level of $w$-mode
excitation in a realistic scenario
\cite{aaks,2ns,kslrev,Tomina99,AP99,ferkdk,ruoff1}.

\section{Gravitational Wave Asteroseismology}

Once gravitational waves are detected the first
task will be to identify the source. This should be possible
from the general character of the waveform and may not require
very accurate theoretical
models, but such models will be of crucial importance
for a deduction of the parameters of the
source. That is, for gravitational-wave ``astronomy''.

The idea behind this presentation is a familiar one in astronomy:
For many years, studies of the light
variation  of variable stars have been used to deduce their internal
structure \cite{unno}.
The Newtonian theory of stellar pulsation was
to a large extent
developed in order to explain the pulsations of Cepheids and RR Lyrae.
This approach, known as Asteroseismology (Helioseismology
in the specific case of the Sun), has been remarkably successful in
recent years. In comparison, the relativistic theory of
stellar pulsation, which has now been developed for thirty years,
has not yet been applied in a similar way.
So far, the relativistic theory has no immediate
connections to observations (that are not already
provided by the Newtonian theory). We believe that this situation
will change once the gravitational-wave
window to the universe is opened, and in this review we discuss how
the information carried by the gravitational-wave signal can be
inverted to estimate the parameters of pulsating stars. That is,
we take a first step towards gravitational-wave asteroseismology.

\subsection{What can we learn from observations?}

Our present understanding of neutron stars comes mainly from
X-ray and radio-timing observations.
These observations provide some insight into the
structure of these objects and the properties of supranuclear
matter. The most commonly and accurately observed parameter
is the  rotation period, and we know that radio pulsars can spin
very fast (the shortest observed period being the 1.56 ms of PSR
1937+21).
Another
basic observable, that can be obtained (in a few cases)
with some accuracy from todays observations,
is the mass of the neutron star.
As Finn \cite{finn} has shown, the observations of radio pulsars
indicate that the mass lies in the range
$1.01 < {M/M_\odot} < 1.64$. Similarly, van der Kerkwijk
et al \cite{Kerk95} find that data for X-ray pulsars  indicate
that $1.04 < {M/M_\odot} < 1.88$. The data used in these two studies
is actually consistent with (if one includes error bars)
$M<1.44 M_\odot$.
We now recall that the various
EOS that have been proposed by theoretical physicists
can be divided into two major categories: i) the ``soft'' EOS which
typically lead to neutron star models with maximum masses around
$1.4 M_\odot$ and radii usually smaller than 10 km, and ii)
  the ``stiff'' EOS
with  maximum values  $M\sim 1.8 M_\odot$ and $R\sim 15$ km
\cite{AB77}.
We thus see that, even though the constraint put on the
neutron star mass by present observations seems strong,
it actually does not rule out many of the proposed EOS.
In order to arrive at a more useful result we are
likely to need detailed observations also of the stellar radius.
Unfortunately, available data provide
little information about the radius. The recent observations of
quasiperiodic oscillations in low mass X-ray binaries indicate
that $R<6M$, but again this is not a severe constraint.
Although a number of attempts have been made, using either X-ray observations
\cite{Lewin} or the limiting spin period of neutron stars \cite{FIP},
to
put constraints on the mass-radius relation,
we do not yet have a method
which can provide the desired answer.
In view of this situation, any method that can be used to infer
neutron star parameters is a welcome addition.
Of specific interest may be the new possibilities
offered once gravitational-wave observations become reality.

Let us suppose that a nearby supernova explodes, say in the Local
Group of galaxies, and is followed by a core collapse that leads
to the formation of a compact object. As the dust from the
collapse settles the compact object pulsates wildly in its various
oscillation modes, generating a gravitational-wave signal which is
composed of an overlapping of different frequencies. We will
assume that the results of Allen et al. \cite{aaks} can be brought
to bear on this situation, i.e. that  most of the energy is
radiated through the $f$-mode, a few $p$-modes and the first
$w$-mode. Our detector picks up this signal, and a subsequent
Fourier analysis of the data stream yields the frequencies and the
energy carried by each mode.

The first question to be answered by the gravitational-wave
astronomer concerns what kind of compact object could produce the
detected signal. Is it a
black hole or a neutron star? The pulsation of  these  objects
lead to qualitatively similar gravitational waves, eg. exponentially
damped oscillations, but the
question should nevertheless be relatively easy to answer.
If more than one of the stellar pulsation modes is observed the answer
is clear, but even if we only observe only one single mode the two
cases should be easy to distinguish.
The fundamental (quadrupole)
quasinormal mode frequency of a Schwarzschild black
hole follows from
\begin{equation}
f \approx 12 {\rm kHz } \left( {M_\odot \over M} \right) ,
\end{equation}
while the associated e-folding time is
\begin{equation}
\tau \approx 0.05 {\rm ms } \left( {M \over M_\odot} \right) \ .
\end{equation}
That is, the oscillations of a 10 $M_\odot$ black hole lie in
the frequency range of the $f$-mode for a typical neutron star.
But the two signals will differ greatly in
 the damping time, the e-folding time of
the black hole being nearly three orders of magnitude shorter than
that of the neutron star $f$-mode.

Having excluded the possibility that our signal came from a
black hole, we want to know  the mass and the radius of
the newly born neutron star. We also want to decide which
of the proposed EOS that best represents
this star. To address these questions we can use a set of empirical
relations  that can be used to
estimate the mass,
the radius and the EOS of the neutron star with good precision.

\subsection{Addressing the inverse problem}

Considering the possibility of a future detection it is relevant
to pose the ``inverse problem'' for gravitational waves from
pulsating stars. Once we have observed the waves, can we deduce
the details of the star from which they originated? To answer this
question we have calculated the frequencies and damping times of
the modes that we expect to lead to the strongest gravitational
waves for a selection of EOS. Nearly twenty years ago Lindblom and
Detweiler \cite{ld} tabulated the frequencies and damping times of
the $f$-mode for a number of EOS. Recently we \cite{ak98} extended
their calculation by
 adding more recent EOS and
providing data also for the $w$- and $p$- modes. These numerical
data were then used to to create useful ``empirical'' relations
between the ``observables'' (frequencies and damping times) and
the parameters of the star (mass, radius and possibly the EOS). We
will now outline how these relations can be used to infer the
stellar parameters from detected mode data.

Let us first consider the frequency of the $f$-mode. It is well
known that the characteristic time-scale of any dynamical process
is related to the mean density ($\bar{\rho}$) of the mass
involved. This notion should be relevant for the fluid oscillation
modes of a star, and  we consequently expect that $\omega_f \sim
\bar{\rho}^{1/2}$. That is, we should normalize the $f$-mode
frequency with the average density of the star. As shown in
\cite{ak98}  the relation between the $f$-mode frequencies and the
mean density is almost linear, and a  linear fitting leads to the
following simple relation:
\begin{equation}
\omega_f \mbox{\rm (kHz)} \approx 0.78 + 1.635 \left(  {{\bar M}
\over {{\bar R}^3}} \right)^{1/2} \ , \label{rfw}
\end{equation}
where we have introduced the dimensionless variables
\begin{equation}
{\bar M} = {M \over {1.4 M_\odot}} \quad \mbox{and} \quad
{\bar R} = {R \over 10 {\rm km} } \ .
\end{equation}
From equation (\ref{rfw}) follows that the typical  $f$-mode frequency
is around 2.4 kHz.

To deduce a similar relation for the damping rate
of the $f$-mode, we can use the rough estimate given by the
 quadrupole formula. That is, the damping time should follow from
\begin{equation}
\tau_f \sim {\mbox{oscillation energy} \over{ \mbox{power emitted in GWs}}}
\sim R\left({R\over M}\right)^3 \ .
\end{equation}
Using this relation and the numerical results for the damping time
of the $f$-mode for various stellar parameters $M$ and $R$, we
find that \cite{ak98}:
\begin{equation}
{1 \over \tau_f ({\rm s})} \approx {{\bar M}^3 \over {\bar R}^4}
\left[ 22.85 - 14.65 \left( {{\bar M}\over {\bar R}} \right)\right]
\label{rftau}
\end{equation}
and it follows that the typical damping time of the $f$-mode is
around 0.15 sec. Analogous empirical relations for the $p$-mode
data are much less robust and useful. This is natural since the
$p$-modes are sensitive to changes in the matter distribution
inside the star (as manifested through changes in the sound
speed).
 In contrast, the
gravitational-wave $w$-modes lead to very robust results.
It is well known
\cite{ks92,akk96} that the $w$-modes do not excite a significant
fluid motion. Thus, they are
more or less independent of the characteristics of the fluid:
The frequencies do not depend on the sound speed and the damping
times cannot be modeled by the quadrupole formula.
Analytic results for model problems
for the $w$-modes \cite{ks86,nils96},
show that the frequency
of the $w$-mode is inversely proportional to the size of the star.
Meanwhile, the
damping time is related to
the compactness of the star, i.e. the more relativistic the star is the longer
the $w$-mode oscillation lasts.
These properties have already been discussed in some detail in \cite{akk96}
for uniform density stars.
One can find
the following relations for the frequency and damping of the first $w$-mode:
\begin{equation}
\omega_w ({\rm kHz}) \approx {1\over {\bar R}}
         \left[ 20.92 - 9.14 \left( {\bar M} \over {\bar R} \right)
\right] \ ,
\label{rww}
\end{equation}
and
\begin{equation}
{1 \over \tau_w ({\rm ms})} \approx {1 \over {\bar M}}
\left[ 5.74 + 103 \left( {{\bar M}\over {\bar R}} \right)
- 67.45 \left( {{\bar M}\over {\bar R}} \right)^2 \right] \ .
\label{rwtau}
\end{equation}
We see that a typical value for the $w$-mode frequency is 11-12 kHz, but
since the frequency depends strongly on the radius of the star
it varies greatly for different EOS.
For example, for a very stiff EOS (L) the
$w$-mode frequency is around 6 kHz while for the
softest  EOS in our set (G) the typical frequency is around 14 kHz.
The $w$-mode damping time is comparable to that of an oscillating
black hole with the same mass,  i.e. it is typically less than a tenth
of a millisecond.

\subsection{Noisy signals}

Suppose that one tries to detect the gravitational waves
associated with the stellar pulsation modes that are excited when
(say) a neutron star forms after a supernova explosion. Since all
modes are relatively short lived, the detection situation is
similar to that for a perturbed rotating black hole
\cite{Ech,finn92}. For each individual mode the signal is
expected to have the following form:
\begin{equation}
h(t)=
\cases{0                                  & for $t<T$, \cr
       {\cal A} e^{- (t-T)/ \tau} \sin[2 \pi f (t-T)] & for $t\geq T$. \cr}
\label{template}
\end{equation}
Here, $\cal A$ is the initial amplitude of the signal, $T$ is
its arrival time, and $f$
and $\tau$ are the frequency and damping time of the oscillation,
respectively.
Since the violent formation of a neutron star is a very complicated
event, the above form of the waves becomes realistic only at the
late stages when the remnant is settling down and its pulsations
can be accurately described as a superposition of the various
modes, either fluid or spacetime ones, that have been excited. At
earlier times ($t<T$) the waves are expected to have a random
character that is completely uncorrelated with the intrinsic noise
of an earth-bound detector. This partly justifies our
simplification of setting the waveform equal to zero for $t<T$.

The energy flux $F$ carried by any weak gravitational wave $h$ is
given by
\begin{equation}
F= {c^3 \over 16 \pi G} | {\dot h} |^2 \;, \label{flux}
\end{equation}
where $c$ is the speed of light and $G$ is Newton's gravitational
constant. Thus, when gravitational waves emitted from a pulsating
neutron star hit such a detector on Earth, their initial amplitude
will be \cite{Schutz97}
\begin{equation}
{\cal A} \sim 2.4 \times 10^{-20} \left( {E_{\rm gw}  \over
10^{-6} M_{\odot} c^2}  \right)^{1/2}  \left( {10 {\rm kpc}  \over
r                       }  \right) \left( {1 {\rm kHz}  \over f }
\right) \left( {1 {\rm ms}   \over \tau                    }
\right)^{1/2} \ , \label{strength}
\end{equation}
where $E_{\rm gw}$ is the energy released through the mode and $r$ is
the distance between detector and source.
In order to dig out this kind of  signal from the noisy output of a
detector one could use templates of the same form as the expected signal
(so called matched filtering).
Following the analysis of Echeverria \cite{Ech} the
signal-to-noise ratio is found to be
\begin{equation}
\left( {S \over N} \right)^2 = \rho^2 \equiv
2 \langle h \mid h \rangle =
{4 Q^2 \over 1+4 Q^2} \; {{\cal A}^2 \tau \over 2 S_n} \;,
\label{SNR}
\end{equation}
with
\begin{equation}
Q \equiv \pi f \tau \;,
\label{Qdef}
\end{equation}
being the quality factor of the oscillation, and
$S_n$ the spectral density of the detector (assumed to be constant
over the bandwidth of the signal).


\subsection{Are the modes detectable?}

Two separate questions must be addressed in any discussion of
gravitational-wave detection. The first one concerns identifying a
weak signal in a noisy detector, thus establishing the presence of
a gravitational wave in the data. The second question regards
extracting the detailed parameters of the signal, e.g., the
frequency and e-folding time of a pulsation mode. To address
either of these issues we need an estimate of the spectral noise
density $S_n$ of the detector.

\begin{figure}
\centering \epsfig{file=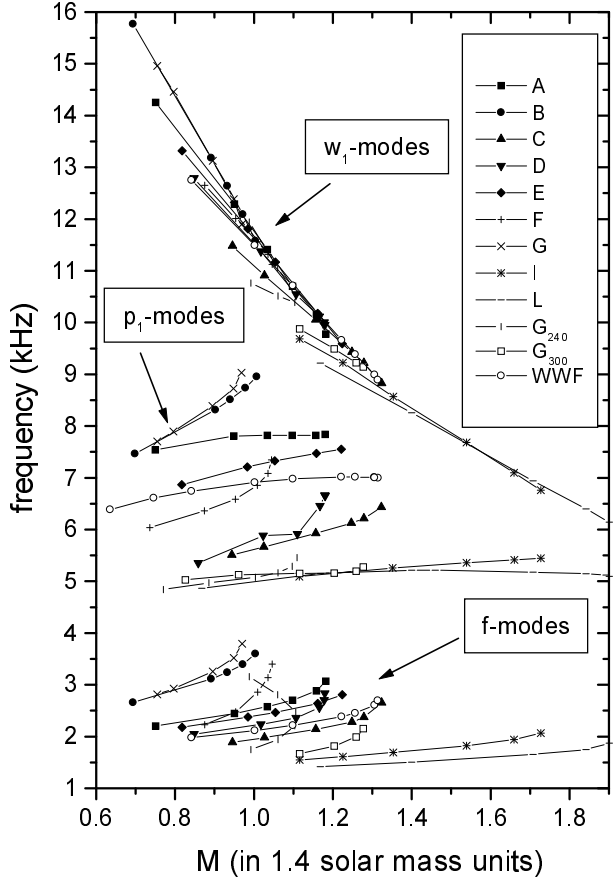, width=10cm} \caption{This
diagram shows the range of mode-frequencies for twelve different
equations of state as functions of the normalized mass (${\bar
M}=M/1.4M_\odot$) of the star. } \label{fig1}
\end{figure}

The pulsation modes of a (non-rotating) neutron star that may be
detectable through the associated gravitational waves all have
rather high frequencies; typically of the order of several kHz. To
illustrate this we show the mode-frequencies for all models
considered in \cite{ak98} as a function of the stellar mass in
Figure~\ref{fig1}. From this figure we immediately see that a
detector must be sensitive to frequencies of the order of 8-12 kHz
and above to observe most $w$-modes. Of course, it is also clear
that some equations of state yield $w$-modes with lower
frequencies. For example, for massive neutron stars with $M\approx
1.8-2.3 M_\odot$ (stiff EOS), which may be needed to account for
data for low-mass X-ray binaries, the $w$-mode frequency could be
as low as 6~kHz (see also recent results for the axial $w$-modes
\cite{BBF99}). The $p$-modes lie mainly in the range 4-8 kHz,
while all $f$-modes have frequencies lower than 4 kHz. This means
that the mode-signals we consider lie in the regime where an
interferometric detector is severely limited by the photon shot
noise. For this reason a detection strategy based on resonant
detectors (bars, spheres or even networks of small resonant
detectors \cite{frasca}) or laser interferometers operating in
dual recycling mode, in order to overcome the overwhelming photon
shot noise at high frequencies \cite{Meers}, seems the most
promising. In fact, the range of mode-frequencies in
Figure~\ref{fig1} provides strong motivation for detailed studies
into the prospects for construction of dedicated ultrahigh
frequency detectors.

\begin{figure}
\centering \epsfig{file=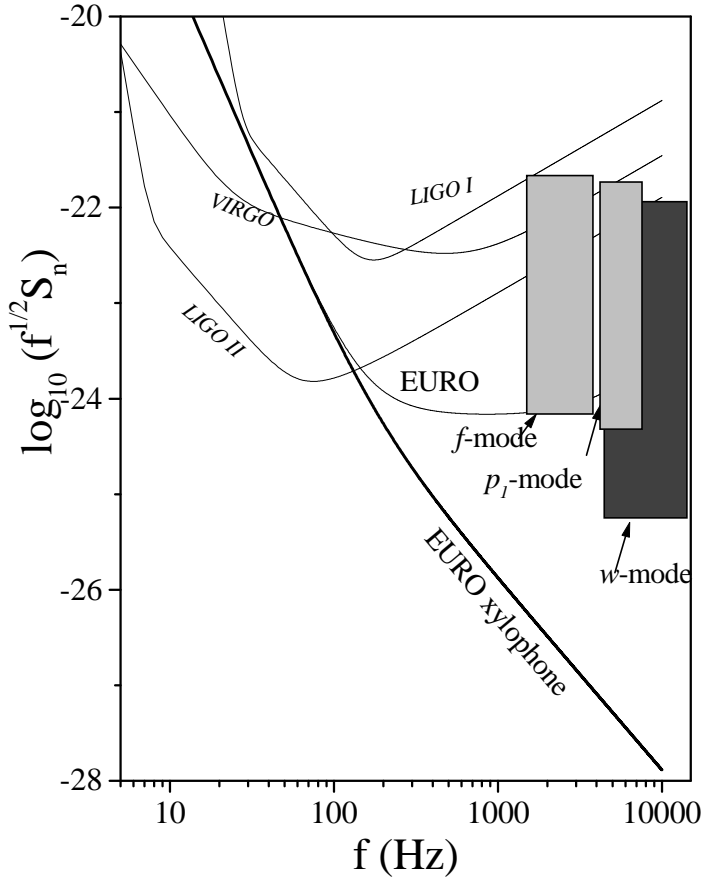,width=10cm} \caption{ The
spectral noise density for the new  generation of laser
interferometric gravitational wave detectors. EURO is a third
generation detector with ideal characteristics for gravitational
wave asteroseismology. The boxes represent the sensitivity needed
to detect a mode into which an energy of $10 ^{-6} M\odot c^2$ is
deposited. The upper limit of each box corresponds to an event in
our galaxy and the lower in the Virgo cluster.} \label{fig2}
\end{figure}

In the following we will compare three different detectors: the
initial and advanced LIGO interferometers, for which
\begin{equation} S_n^{1/2} \approx h_m \left( { f \over \alpha
f_m} \right)^{3/2} {1\over \sqrt{f}}\ \mbox{Hz}^{-1/2} \;,
\end{equation}
with $h_m=3.1\times10^{-22}$, $\alpha=1.4$ and $f_m=160$~Hz for
the initial configuration, and $h_m=1.4\times10^{-23}$,
$\alpha=1.6$ and $f_m=68$~Hz for the advanced configuration
\cite{flanagan}. We also consider an ``ideal'' detector that is
tuned to the frequency of the mode and has sensitivity of the
order of $S_n^{1/2} \approx 10^{-24}$ Hz$^{-1/2}$. This is the
sensitivity goal of the new generation of detectors under
construction, see Figure~ \ref{fig2}. As an example of a suitably
advanced instrument we will take the so called EURO detector, for
which the noise-level curves have been estimated by Sathyaprakash
and Schutz (private communication). It should be noted that the
Advanced LIGO estimates are roughly valid also for spherical
detectors such as TIGA, cf. Harry, Stevenson and Paik
\cite{harry}.

The detectability of the $f$, $p$ and $w$-modes for different
detectors can be assessed from (\ref{SNR}). The main problem in
doing this is the lack of realistic simulations providing
information about the level of excitation of various modes in
an astrophysical situation. Still, given the frequency and damping
rate of a specific mode we can ask what amount of energy must be
channeled through the mode in order for it to be detectable by a
given detector. We immediately find that
detection of pulsating neutron stars from outside our own galaxy
is very unlikely. Let us consider a ``typical'' stellar model for
which the $f$-mode has parameters $f_f=2.2$~kHz and $\tau_f=0.15$~s,
this corresponds to a $1.4M_\odot$ neutron star according to the
Bethe-Johnson equation of state \cite{ak98}.
For this example we find that the $f$-mode in the
Virgo cluster (at 15 Mpc) must carry an energy equivalent to more
than $0.3 M_\odot c^2$ to lead to a signal-to-noise ratio of 10 in
our ideal detector. Given that the total energy estimated to be
radiated as gravitational waves in a supernova is at the level of
$10^{-5}-10^{-6} M_\odot c^2$, we cannot realistically expect to
observe mode-signals from far beyond our own galaxy.

This means that the number of detectable events may be rather low.
Certainly, one would not expect to see a supernova in our galaxy
more often than once every thirty years or so. Still, there are a
large number of neutron stars in our galaxy, all of which may be
be involved in dramatic events (see the introduction for some
possible scenarios) that lead to the excitation of pulsation modes.
The energies required to make each mode detectable (with a
signal-to-noise ratio of 10) from a source at the center of our galaxy (at 10
kpc) are listed in Table~\ref{tab1}. In the table we have used the
data for the ``typical'' stellar model, for which the characteristics
of the $f$-mode were given
above, $f_p=6$~kHz and $\tau_p=2$~s, and $f_w=11$~kHz and
$\tau_w=0.02$~ms. This data indicates that, even though the event
that excites the modes must be violent, the energy required to
make each mode detectable is not at all unrealistic. In fact, the
energy levels required for both the $f$- and $p$-modes are such
that detection of violent events in the life of a neutron star
should be possible, given the Advanced LIGO detectors (or
alternatively spheres with the sensitivity proposed for TIGA). On
the other hand, detection of $w$-modes with the broad band
configuration of LIGO seems unlikely.
Detection of these modes,
which would correspond to observing a purely relativistic
phenomenon, requires dedicated high frequency detectors operating
in the frequency range above 6 kHz. We believe that the data in
Table~\ref{tab1} illustrates that neutron star pulsation modes may
well be detectable from within our galaxy. The first
detection may come as soon as the first generation of LIGO
detectors come on line, but it may be more realistic
to expect that we need a third generation detector (such as EURO)
to truly probe the pulsation modes of neutron stars.

\begin{table}
\begin{center}
\begin{tabular}{*{4}{r}}
\multicolumn{4}{l}{}\\
\hline \\
 Detector  & $f$-mode & $p$-mode & $w$-mode \\[0.5ex]
\hline
\\[0.5ex]
LIGO I  & $4.9\times10^{-5}$ & $4.0\times 10^{-3}$ &  $6.8\times 10^{-2}$ \\[0.5ex]
LIGO II & $8.7\times10^{-7}$ & $7.0\times 10^{-5}$ & $1.2\times 10^{-3}$ \\[0.5ex]
Ideal   & $1.4\times10^{-8}$ & $1.3\times 10^{-7}$ &  $6.4\times 10^{-7}$ \\[0.5ex]
\hline
\end{tabular}
\vspace{3mm}
\caption{The estimated energy (in units of $M_\odot c^2$)
  required in each mode in order to lead to a
  detection with signal-to-noise ratio of 10 from a pulsating neutron star at
  the center of our galaxy (10 kpc), cf. Eqs (3,4). The given data
  corresponds to a $1.4M_\odot$ star with the Bethe-Johnson equation
  of state.}
\label{tab1}
\end{center}
\end{table}
\subsection{How well can we determine the mode parameters?}

Let us now discuss the precision with which we can hope to infer
the details of each pulsation mode.
We can compute the relative measurement error in
the frequency and the damping time of the waves by some
appropriately designed detector \cite{KAA01}.
After introducing a convenient
parameter ${\cal P}$, defined by
\begin{equation}
{\cal P}^{-1} = \left( {S_{n}^{1/2}   \over 10^{-24}  {\rm
Hz^{-1/2}}} \right) \left( {r             \over 10  {\rm kpc} }
\right) \left( {E_{\rm gw}    \over 10^{-6}   M_{\odot} c^2}
\right)^{-1/2} \;,
\end{equation}
we find that the error estimates take the following form
\begin{equation}
{\sigma_f \over f} \simeq 0.0042 \  {\cal P}^{-1} \ {\sqrt
{1-2Q^2+8Q^4 \over 4Q^4}} \ \left( {\tau           \over 1  {\rm
ms}           }   \right)^{-1}   , \label{relerrorf}
\end{equation}
and
\begin{equation}
{\sigma_\tau \over \tau} \simeq 0.013 \ {\cal P}^{-1} \ {\sqrt
{10+8Q^2  \over Q^2}}    \ \left(  {f            \over 1  {\rm
kHz}           }   \right)        . \label{relerrortau}
\end{equation}
Also, for the time of arrival of the gravitational wave signal we
get from
\begin{equation}
\sigma_T \simeq 0.0042 \ {\cal P}^{-1} {\rm ms}         \ .
\label{relerrortime}
\end{equation}
To illustrate these results we list  in Table~\ref{tab2}
the relative errors associated
with the parameter extraction for the ``typical'' $1.4M_\odot$
stellar model we used in the previous section.
We assume that each
mode carries the energy required for it to be observed with
signal-to-noise ratio of 10, cf.  Table~\ref{tab1}. (This is a
convenient measure since it is independent of the particulars of
the detector.)

\begin{table}
\begin{center}
\begin{tabular}{*{4}{l}}
\multicolumn{4}{l}{}\\
\hline
Mode & $\sigma_f/f$ & $\sigma_\tau/\tau$ & $\sigma_T$~($10^{-6}$~s) \\[0.5ex]
\hline
\\[0.5ex]
$f$-mode & $8\times10^{-6}$ & $0.02$ & 1.1 \\[0.5ex]
$p$-mode & $3\times10^{-7}$ & $0.02$ & 0.38 \\[0.5ex]
$w$-mode & $0.01$ & $0.03$ & 0.177 \\[0.5ex]
\hline
\end{tabular}
\vspace{3mm}
\caption{The relative errors in the extraction of the mode
  parameters assuming a signal-to-noise ratio of 10.
The relatively large error in estimating $f_{w}$
is due to the rapid damping of these modes (small $Q$ factor),
cf. Eq.~(12). In other words,
we will not be able to detect more than a couple of cycles of a
  $w$-mode whereas many hundred cycles of both the $f$- and
  the $p$-modes could be observed. }
\label{tab2}
\end{center}
\end{table}

From the sample data in Table~\ref{tab2} one sees clearly that,
while an extremely accurate determination of the frequencies of both the $f$-
and the $p$-mode is possible, it would be much harder to infer
their respective damping rates. It is also clear that an accurate
determination of both the $w$-mode frequency and damping will be
difficult. To illustrate this result in a different way, we can
ask how much energy must be channeled through each mode in order
to lead to a 1\%
relative error in the frequency or the damping
rate, respectively. Let us call the corresponding energies
$E_f$ and $E_\tau$.
This measure will then be detector dependent, so we list the
relevant estimates for the three detector configurations used
in Table~\ref{tab1}. When the data is viewed in this way,
cf. Table~\ref{tab3},
we see that an accurate extraction of $w$-mode data will not be
possible unless a large amount of energy is released through these
modes. Furthermore, one would clearly need a detector that is
sensitive at ultrahigh frequencies.

\begin{table}
\begin{center}
\begin{tabular}{*{7}{l}}
\multicolumn{7}{l}{}\\
\hline
 Detector &   $f$-mode & &  $p$-mode & & $w$-mode & \\[0.5ex]
&  $E_f$ & $E_\tau$   & $E_f$ & $E_\tau$   & $E_f$ & $E_\tau$ \\[0.5ex]
\hline
 LIGO I & $3\times10^{-9}$ & $2\times 10^{-2}$ & $3\times 10^{-10}$  & --- & --- & --- \\[0.5ex]
 LIGO II  & $5\times10^{-11}$ & $3\times 10^{-4}$ & $5\times 10^{-12}$ & $3\times10^{-2}$ & $0.2$
& --- \\[0.5ex]
 Ideal & $9\times10^{-13}$ & $6\times 10^{-6}$ & $9\times
10^{-15}$& $5\times10^{-5}$ & $9\times 10^{-5}$ & $6\times
10^{-4}$ \\[0.5ex] \hline
\end{tabular}
\caption{The estimated energy (in units of $M_\odot c^2$)
  required in each mode in order to lead to a relative error of 1\%
  in the inferred mode-frequency ($E_f$) and damping rate
  ($E_\tau$). In cases where no entry is given, the required energy is
  unrealistically high (typically larger than $M_\odot c^2$).
  The distance to the source is assumed to be 10 kpc.}
\label{tab3}
\end{center}
\end{table}

\subsection{Revealing the position of the source}

In the previous section we discussed issues regarding the
detectability of a mode-signal, and the accuracy with which the
parameters of the mode could be inferred from noisy gravitational
wave data. Let us now assume that we have detected the mode and
extracted the relevant parameters. We then naturally want to constrain
the supranuclear equation of state by deducing the
mass and the radius of the star. In
principle, the mass and the radius can be deduced from any two
observables, \cite{ak98}. In the absence of detector noise, several
combinations look promising, but in reality only few
combinations are likely to be useful.

As with other kinds of gravitational-wave sources, a network of at
least three detectors is needed to pinpoint the location of the
source in the sky. The difference in arrival time for the three
 detectors could be used to determine the position of the
source. The higher the accuracy in measuring the time of arrival
at each detector, the more precise will be the positioning of the
source. Two remote detectors, at a distance $d$ apart from each
other will receive the signal with a temporal difference of
\begin{equation}
\Delta T={d \over c} \cos\theta \;,
\label{deltaT}
\end{equation}
where $c$ is the speed of light, and $\theta$ is the angle between
the line joining the two detectors and the line of sight of the source.
Therefore, the accuracy by which this angle can be measured is
\begin{equation}
\Delta \theta={\sqrt{2} \sigma_{T} c \over d \sin\theta}\;.
\label{deltatheta}
\end{equation}
The $\sqrt{2}$ arises from the measurement errors of the two times
of arrival. If one assumes an `L' shaped network of 3 detectors
with arm length of $d= 10,000$~km, Eqs.~(\ref{relerrortime}) and
(\ref{deltatheta}) lead to an error box on the sky with angular
sides of $1^{\rm o}$, at most (for specific areas of the sky, and large
signal-to-noise ratios the angular sizes could be much smaller).
This is quite interesting since one could then correlate
the detection of gravitational waves with radio, X-ray or gamma-ray
observations directed towards that specific corner of the sky.

\section{The unstable $r$-modes}

So far we have discussed the modes of (essentially)
non-rotating stars. A more optimistic scenario is based on the notion
that various modes of oscillation may be unstable in a rotating
star. Following the  serendipitous discovery of such an
instability in the so-called
$r$-modes \cite{a97,FM98}, this area of research has
attracted considerable attention.
Should such an instability operate in a young neutron star
it may lead to the emission of copious amounts
of gravitational waves \cite{owen}.
Such gravitational waves have been estimated
to be detectable
for sources in the Virgo cluster (at 15-20~Mpc).
If we suppose that most
newly born neutron stars pass through a phase where this
kind of instability is active, several such events  should be observed per year
once the advanced interferometers come into operation. This is a very
exciting prospect, indeed.

In this review article we give a brief introduction to these
recent ideas  and suggestions. For an exhaustive
discussion we refer the reader to \cite{ak01}.

The $r$-modes are unstable to the emission of gravitational
waves via a mechanism that was first suggested by
Chandrasekhar \cite{chandra}. Subsequent work by
Friedman and Schutz \cite{FS78} showed that this instability is a generic
feature of all rotating fluids, and in the case of $r$-modes
Friedman and Morsink \cite{FM98} showed that
that the new instability is generic for toroidal perturbations of
relativistic stars.

In a simple description, the instability
works as follows:
In a rapidly rotating star a backward moving mode (as measured by a
co-rotating observer)
can be dragged forward according to an inertial observer. This means that
  the mode radiates positive angular momentum, even though
the angular momentum of the mode remains negative because the
perturbed star has lower net angular momentum than the unperturbed
star. As positive angular momentum is removed from the mode, its
angular momentum becomes increasingly negative, implying that its
amplitude increases. The mode grows due to the emission of
gravitation radiation. In other words, a mode is unstable if it is
prograde relative to infinity and retrograde relative to the star.
For many years the investigations into the relevance of the CFS
instability was focussed on the spheroidal $f$-mode (for a recent
review, see \cite{Nikos}). The reason for this is that this mode
was considered the most important one as far as gravitational
waves are concerned. Hence, it came as some surprise that the
instability associated with the toroidal $r$-modes is considerably
stronger than the $f$-mode one.

In fact, toroidal modes of relativistic stars attracted little
attention until recently. Such perturbations of non-rotating stars
lead to a set of zero frequency modes in Newtonian theory,
complemented by gravitational-wave modes ($w$-modes) in the
relativistic description \cite{CF91,kdk94}. Toroidal modes of a
Newtonian star describe stationary  horizontal fluid currents, and
do not induce variations  in the density and pressure. When the
star is set into rotation the Coriolis force provides a weak
restoring force that gives the toroidal modes true dynamics, and
the fluid undergoes oscillations with a frequency (measured by an
inertial observer at infinity)
\begin{equation}
\sigma = -m\Omega + {2m \Omega \over {\ell(\ell+1)}} \ ,
\label{eq_rmode}
\end{equation}
where $\Omega$ is the rotational frequency of the star and $\ell$ and $m$
are the spherical harmonic indices. This result, that identifies the $r$-modes,
follows from a Newtonian
treatment of oscillations of slowly rotating stars.
As one can easily deduce from their frequency the $r$-modes are always
retrograde in the corotating frame
and their phase velocity is always smaller than that of the
rotation of the star; thus they are generically unstable
independently of the rotation rate of the star.

\begin{figure}
\centering \epsfig{file=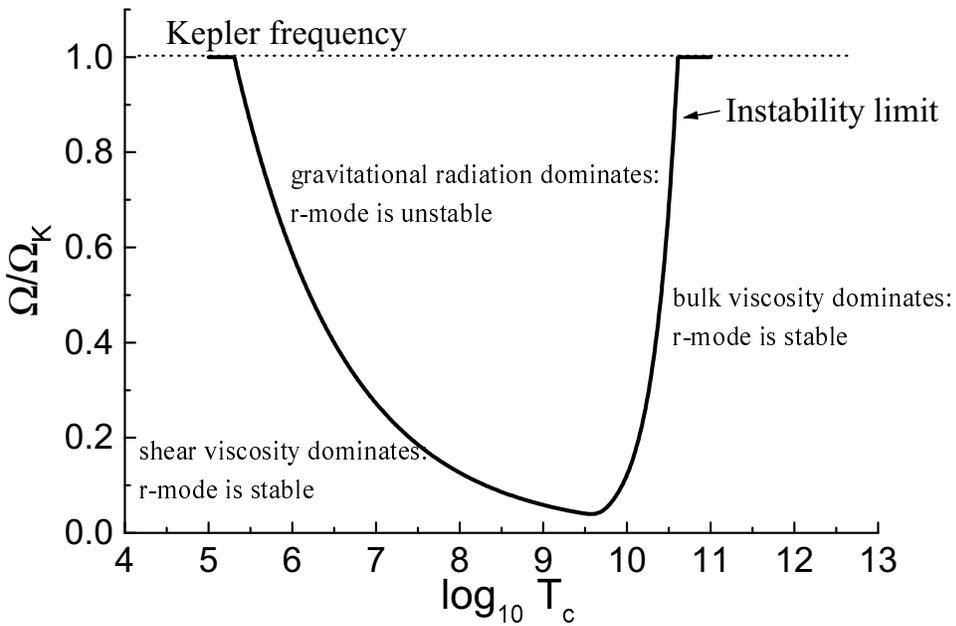,width=13cm} \caption{The
critical rotation rate at which shear viscosity (at low
temperatures) and bulk viscosity (at high temperatures) balance
the $r$-mode instability} \label{fig3}
\end{figure}

That a mode is formally unstable in a perfect fluid star does not
in itself mean that it will be allowed to grow and affect, for
example, the star's spin evolution. In any ``realistic'' star
there are dissipation mechanisms that may halt growth of an
instability. In the simplest model, one must account for the
effects of bulk and shear viscosity. Comparison between the
damping times due to viscosity and the growth time due to
gravitational radiation provides a criterion for the significance
of the instability. The damping/growth time associated with each
mechanism can be estimated from the ratio of the energy loss to
the available mode-energy (measured in the rotating frame) i.e.
\begin{equation}
{1 \over \tau_{\rm diss}} = - {{\dot E}_{\rm diss}\over 2E_{\rm mode}}
\end{equation}
In  our case the onset of the instability is signalled by
\begin{equation}
 {1 \over \tau} = -{1 \over \tau_{gw}}
+ {1\over \tau_{sv}} +{1\over \tau_{bv}}  =0.
\end{equation}
In this relation, the growth time due to gravitational waves {\em
(gw)} is temperature independent while both bulk {\em (bv)} and
shear viscosity {\em (sv)} are strongly dependent on the internal
temperature of the star. We can now find the critical rotation
frequency at which the $r$-mode becomes unstable for the relevant
values of the core temperature of the star. Detailed calculations
have shown that bulk viscosity damps any fluid motion for
temperatures higher than $10^{10}$ K, while shear viscosity
dominates for temperatures below $10^6$ K. This means that there
is a ``window of opportunity'' between $10^6 - 10^{10}$  K where
the $r$-mode instability is active and may play an astrophysical
role \cite{lom98,aks99}. This instability window is shown in
Figure~\ref{fig3}.

Finally, it is worth pointing out that the gravitational
radiation emitted from the $r$-modes
comes primarily from the time-dependent {\em mass-currents}.
This is the gravitational analogue of magnetic multipole radiation
and the $r$-mode  instability is unique among expected
astrophysical sources of gravitational radiation in radiating primarily by
gravitomagnetic effects. The detectability of these gravitational
waves is discussed in detail in \cite{ak01}.

\subsection{A novel scenario}

Before we conclude this review, it is appropriate to comment on
the fact that we neglected  the effects of both rotation and
magnetic fields on the pulsation modes in our discussion of
asteroseismology. This was (obviously!) not because rotation and
magnetic fields play an insignificant role. On the contrary, we
expect rotation to be highly relevant in many cases (in particular
since various instabilities may set in above a critical spin
rate), and the recent evidence for magnetars make it clear that
one cannot neglect the role of the magnetic field.  Still, as far
as most pulsation modes are concerned, one would expect rotation
to have a significant effect only for neutron stars with very
short period, and the present study may well be reasonable for
stars with periods longer than, say, 20 ms. Furthermore, it has
been argued that neutron stars are typically born slowly rotating
\cite{spruitt}. If that is the typical case, then our results
could be relevant also for most newly born neutron stars. A more
pragmatic reason for not including rotation in the present study
is that detailed data for modes of rotating neutron stars is not
yet available. Once such data has been computed the present study
can be extended to incorporate rotational effects. There are
currently efforts by various groups around the world to estimate
the effects of rotation on $f$-, $p$- and $w$-modes (and also
calculate the $r$-modes) in the framework of general relativity.
There are already preliminary results for slowly-rotating stars
\cite{kojima93,a97} but a complete study of both slowly and fast
rotating stars is outstanding.

One of the main rotational effects is the splitting of the various
frequencies, i.e. the frequency is no longer degenerate with
respect to the harmonic index $m$. For example, the quadrupole
$f$-mode oscillations ($l=2$) will emit gravitational waves with
all five possible values of $m$ (-2,-1,0,1,2). This splitting of
the spectrum will create significant problems in using the
empirical relations such as (\ref{rfw}). One has to derive a
similar formula with an additional term of the form $m\Omega
(M/R^3)^{1/2}$ in order to include in an appropriate way the
rotational corrections. In this case the prior knowledge of the
rotation rate of the star may be needed. Potentially, this
information can be deduced from  gravitational waves due to the
$r$-mode instability. In this case the frequency is directly
proportional to the rotation rate of the star, cf. equation
(\ref{eq_rmode}), and the $r$-mode instability signal is expected
to be much stronger than those of the other modes. Hence, one can
immediately infer the rotation of the star. The rotation rate will
obviously change due to the $r$-mode, but for time intervals of
the order of 1-2 secs (the typical damping time of other modes) it
can be considered as constant. The information about the rotation
rate can then be combined with empirical formulas that can be
derived for the modes of rotating stars in order to extract all of
the parameters of the star.

Strong magnetic fields can also affect the mode frequencies of the
neutron stars. Preliminary results for Newtonian stars have shown
that typical magnetic fields of the order of $10^{11}-10^{14}$
Gauss, do not produce significant shifts in the mode frequencies.
One must to consider extremely strong magnetic fields, like in
magnetars ($10^{15}-10^{16}$~G) in order to get a significant
frequency shift. In fact, this highlights an interesting issue for
the magnetars: if one can detect gravitational waves from the
starquakes induced by their strong magnetic fields, a possible
shift in the spectrum may provide a measure of the
 magnetic field strength.

Finally, it is worth mentioning that a detection mode
pulsation modes from old, cold neutron stars could also provide
unique insights into the superfluid nature of neutron star cores.
It is  generally
believed that once a neutron star cools below a
few times $10^9$~K (a few months after its birth) the bulk of its
core will become superfluid. Thus, the more than 1000 observed
pulsars provide useful laboratories for studying
large scale superfluidity.
In the simplest description, a superfluid neutron star core can be
discussed in terms of two distinct fluids. One of these fluids
represents the superfluid neutrons and the other fluid represents
all charged components (which
are expected to be coupled on a
relatively short timescale). The fact that these two
fluids --- the ``neutrons'' and the
``protons'' --- can flow more or less independently provides one of the
main distinguishing
dynamical features of a superfluid neutron star.
In particular, one can show that two sets of pulsation modes
are interlaced in the spectrum of a superfluid neutron star core
\cite{Comer1,Comer2}.
One set of modes are the familiar $p$-modes, for which the two fluids
tend to move together. The other set of modes  are
distinguished by the fact that the protons and neutrons are largely
``countermoving''. This class of modes is unique to the two-fluid system.
Of particular interest for our current discussion is the fact that
the ``superfluid mode'' frequencies are (locally) approximated by
\begin{equation}
\omega^2_s \approx {m_{p} \over m^*_p} {l (l + 1) \over r^2} c^2_p \ ,
\end{equation}
where $c^2_p$ is (roughly) the sound speed in the proton fluid, $r$
is the radial coordinate, and $l$ is the index of the relevant
spherical harmonic $Y_{lm}(\theta,\varphi)$ used to describe the
angular dependency of the mode.
From this relation it is clear that
an observation of these modes would provide potentially  unique
information regarding the nature of large scale superfluidity,
and could put useful constraints on  crucial parameters
such as the
ratio between the
``bare'' and ``effective'' proton masses $m_p$ and $m_p^*$
(estimates to lie in the range $0.3 \le  {m_p^* / m_p} \le 0.8$).
We think this is a very exciting prospect that
should motivate future efforts in this field.

\section*{Acknowledgements}
We appreciated useful comments by Th. Apostolatos, J. Ruoff and N. Stergioulas.
This work has been supported by the EU Network
Contract No. HPRN-CT-2000-00137


\section*{References}
\begin{thebibliography}{99}

\bibitem{ak96}
        Andersson N.,  Kokkotas  K.D., (1996), Phys.~Rev.~Lett., {\bf 77}, 4134

\bibitem{ak98}
        Andersson N.,  Kokkotas  K.D., (1998), MNRAS, {\bf 299}, 1059

\bibitem{KAA01}
        Kokkotas K.D., Apostholatos Th., Andersson N. (2001) MNRAS {\bf 320},
        307
\bibitem{BBF99}
        Benhar O., Berti E., Ferrari V., (1999), MNRAS, {\bf 310}, 797

\bibitem{Yip99} Yip C.W., Chu M.-C., Leung P.T. (1999), ApJ, {\bf 513}, 849

\bibitem{duncan}
        Duncan R.C., (1998) ApJ, {\bf 498}, L45

\bibitem{ak01}
        Andersson N.,  Kokkotas  K.D., (2001), Int. J. Mod. Phys. D
    , {\bf 10}, 381

\bibitem{ks92}
    Kokkotas K.D.,  Schutz B.F. (1992) MNRAS {\bf 255}, 119

\bibitem{kokkotas97}
        Kokkotas K.D. (1997) {\em "Pulsating relativistic stars"} in {\em Relativistic Gravitation
        and Gravitational Radiation}, ed. by J.-A. Marck and J.-P. Lasota
        (Cambridge University Press), Cambridge, pp.89

\bibitem{kslrev} Kokkotas K.D., Schmidt B.G. (1999)  Living Reviews in Relativity,
    1999-2: \\
http://www.livingreviews.org/Articles/Volume2/1999-2kokkotas

\bibitem{aaks}
        Allen G., Andersson N.,  Kokkotas  K.D., Schutz B.F, (1998),
        Phys.Rev.D, {\bf 58}, 124012

\bibitem{2ns}
        Allen G., Andersson N, Kokkotas K.D., Laguna P., Pullin J.A.,
        Ruoff J. (1999),
        Phys. Rev. D, {\bf 60}, 104021

\bibitem{ruoff} Ruoff J.,  (2001) Phys. Rev. D,{\bf 63}, 064018

\bibitem{Tomina99} Tominaga K., Saijo M., Maeda K., (1999) Phys.~Rev. D {\bf 60}, 24004

\bibitem{AP99} Andrade Z., Price R.H., (1999), Phys. Rev. D {\bf 60}, 104037

\bibitem{ferkdk} Ferrari V.,  Kokkotas K.D., (2000) Phys. Rev. D {\bf 62}, 107504

\bibitem{ruoff1} Ruoff J., Laguna P., Pullin J., (2001) Phys. Rev. D {\bf 63}, 064019

\bibitem{blaes}
       Blaes O., Blandford R., Goldreich P., Madau P., (1989), ApJ {\bf 343}, 839

\bibitem{mock}  Mock P.C., Joss P.C., (1998), ApJ, {\bf 500}, 374

\bibitem{dt92} Duncan R.C., Thompson C. (1992), ApJ, {\bf 392}, L9

\bibitem{baum} Baumgarte T.W., Janka H-T., Keil W., Shapiro S.L., Teukolsky S.A.
        (1996), ApJ, {\bf 468}, 823

\bibitem{KS95}   Kokkotas K.D., Sch\"afer G., (1995), M.N.R.A.S., {\bf 275}, 301

\bibitem{unno} Unno W., Osaki Y., Ando H., Shibahashi H., (1989) {\em Nonradial
        oscillations of stars}, University of Tokyo Press

\bibitem{cow41}    T.G. Cowling T.G., (1941) MNRAS, {\bf 101}, 367

\bibitem{mcdermott} McDermott P.N.,  van Horn H.M.,  Hansen C.J. (1988) ApJ, {\bf 325}, 725

\bibitem{strohmayer} Strohmayer T.E. (1991) ApJ {\bf 372},573

\bibitem{ks86}    Kokkotas K.D,  Schutz B.F., (1986) Gen. Rel. Grav. {\bf 18} 913

\bibitem{kojima88} Kojima Y., (1988) Prog. Theor. Phys. {\bf 79}, 665

\bibitem{akk96} Andersson N., Kojima Y., Kokkotas K.D. (1996)
    {\bf 462}, 855

\bibitem{finn}  Finn L.S. (1994) Phys. Rev. Lett, {\bf 73}, 1878

\bibitem{Kerk95} van Kerkwijk M.H., van Paradijs J., Zuiderwijk E.J. (1995) A\&A {\bf 303}, 497

\bibitem{AB77} Arnett W.D., Bowers R.L. (1977) ApJS, {\bf 33}, 415

\bibitem{Lewin} Lewin W.H.G., van Paradijs J., Taam R.E., (1993) Space Sci. Rev. {\bf 62}, 223

\bibitem{FIP} Friedman J.L., Ipser J.R., Parker L., (1986) ApJ {\bf 305}, 115

\bibitem{ld} Lindblom L., Detweiler S., (1983), Ap.J.Suppl.  {\bf 53}, 73

\bibitem{nils96} Andersson N., (1996) Gen. Relativ. Gravitation {\bf 28}, 1433

\bibitem{Ech} Echeverria F., (1989) Phys. Rev. D. {\bf 40}, 3194

\bibitem{finn92} Finn L.S. (1992) Phys. Rev. D, {\bf 46} 5236

\bibitem{Schutz97}
        Schutz B.F, (1997) {\em ''The detection of gravitational waves''} in {\em Relativistic Gravitation
        and Gravitational Radiation} ed. by J.-A. Marck J.-A., J.-P. Lasota
        (Cambridge University Press), Cambridge, pp. 447

\bibitem{frasca} Frasca S., Papa M.A., (1995) Int. J. Mod. Phys., {\bf 4}, 1

\bibitem{Meers} Meers B.J., (1988) Phys. Rev. D {\bf 38}, 2317

\bibitem{flanagan} Flanagan E.E., Hudges S. (1998), Phys. Rev. D, {\bf 57}, 4235

\bibitem{harry} Harry G.M., Stevenson T.R., Paik H.J., (1996) Phys. Rev. D, {\bf 54}, 2409

\bibitem{a97}   Andersson N., (1998), ApJ, {\bf 502}, 708

\bibitem{FM98}   Friedman J.L., Morsink S. (1998), ApJ, {\bf 502}, 714

\bibitem{owen}
        B.J.~Owen, L.~Lindblom, C.~Cutler, B.F.~Schutz, A.~Vecchio,
        N.~Andersson, (1998) Phys. Rev. D {\bf 58}, 084020

\bibitem{chandra} Chandrasekhar S., (1970) Phys. Rev. Lett. {\bf 24}, 611

\bibitem{FS78} Friedman J.F.,  Schutz B.F., (1978) Ap.J. {\bf 222}, 281

\bibitem{Nikos} Stergioulas N., (1998) Living Reviews in Relativity,
    1998-8  \\
    http://www.livingreviews.org/Articles/Volume1/1998-8stergio/

\bibitem{CF91} Chandrasekhar S., Ferrari V., (1991) Proc.R.Soc. London Ser A, {\bf 433}, 423

\bibitem{kdk94} Kokkotas K.D., (1994) MNRAS, {\bf 268}, 1015

\bibitem{lom98} Lindblom L., Owen B.J., Morsink S.M., (1998) Phys. Rev. Lett. {\bf 80} 4843

\bibitem{aks99} Andersson N.,  Kokkotas K.D., Schutz B.F., (1999) Ap.J. {\bf 510}, 846

\bibitem{spruitt} Spruitt H., Phinney E.S., (1998) Nature {\bf 393}, 139

\bibitem{kojima93} Kojima Y., (1993) ApJ, {\bf 414}, 247

\bibitem{Comer1} Comer G.L., Langlois D., Lin L.M. (1999) Phys. Rev. D {\bf 60}, 104025

\bibitem{Comer2} Andersson N., Comer  G.L. (2001) MNRAS in
press, astro-ph/0101193

\end{thebibliography}
\end{document}